\documentclass[a4paper,fleqn,usenatbib]{mnras}

\usepackage{newtxtext,newtxmath}

\usepackage[T1]{fontenc}
\usepackage{ae,aecompl}

\usepackage{graphicx}	
\usepackage{amsmath}	
\usepackage{amssymb}	

\usepackage{multirow}

\hypersetup{draft}

\newcommand{\RM}{R_\mathrm{MOA}}
\newcommand{\VM}{V_\mathrm{MOA}}

\title[ob161195]{The Lowest Mass Ratio Planetary Microlens: OGLE~2016--BLG--1195Lb}

\author[I. A. Bond et al.]{
I.~A. Bond,$^{1}$\thanks{E-mail: i.a.bond@massey.ac.nz (IAB)}
D.~P. Bennett,$^{2}$
T. Sumi,$^{3}$
A. Udalski,$^{4}$
D. Suzuki,$^{2}$
\newauthor
N.~J. Rattenbury,$^{5}$
V. Bozza,$^{6,7}$
N. Koshimoto,$^{3}$
F. Abe,$^{8}$
Y. Asakura,$^{8}$
\newauthor
R.~K. Barry,$^{2}$
A. Bhattacharya,$^{2,9}$
M. Donachie,$^{5}$
P. Evans,$^{5}$
A. Fukui,$^{10}$
\newauthor
Y. Hirao,$^{3}$
Y. Itow,$^{8}$
M.~C.~A. Li,$^{5}$
C.~H. Ling,$^{1}$
K. Masuda,$^{8}$
Y. Matsubara,$^{8}$
\newauthor
Y. Muraki,$^{8}$
M. Nagakane,$^{3}$
K. Ohnishi,$^{11}$
C. Ranc,$^{2}$
To. Saito,$^{12}$
A. Sharan,$^{5}$
\newauthor
D.~J. Sullivan,$^{13}$
P.~J. Tristram,$^{14}$
T. Yamada,$^{3}$
T. Yamada,$^{15}$
A. Yonehara,$^{15}$
\newauthor
J. Skowron$^4$,
M.K. Szyma{\'n}ski$^4$,
R. Poleski$^{4,16}$,
P. Mr{\'o}z$^{4}$,
I. Soszy{\'n}ski$^4$,
\newauthor
P. Pietrukowicz$^4$,
S. Koz{\l}owski$^4$,
K. Ulaczyk$^{4,17}$,
and M. Pawlak$^4$
\\
$^{1}$Institute for Natural and Mathematical Sciences, Massey University, Private Bag 102904 North Shore Mail Centre,\\Auckland 0745, New Zealand\\
$^{2}$Laboratory for Exoplanets and Stellar Astrophysics, NASA/Goddard Space Flight Center, Greenbelt, MD 20771, USA\\
$^{3}$Department of Earth and Space Science, Graduate School of Science, Osaka University, 1-1 Machikaneyama, Toyonake,\\ Osaka 560-0043, Japan\\
$^{4}$Warsaw University Observatory, Al. Ujazdowskie 4, PL-00-478 Warszawa, Poland\\
$^{5}$Department of Physics, University of Auckland, Private Bag 92019, Auckland, New Zealand\\
$^{6}$Dipartimento di Fisica ``E. R. Caianiello'', Universit{\`a}  di Salerno, Via Giovanni Paolo II, 84084 Fisciano (SA), Italy\\
$^{7}$Istituto Nazionale di Fisica Nucleare, Sezione di Napoli, Via Cintia, 80126, Napoli, Italy\\
$^{8}$Institute of Space-Earth Environmental Research, Nagoya University, Furo-cho, Chikusa, Nagoya, Aichi 464-8601, Japan\\
$^{9}$Department of Physics, University of Notre Dame, Notre Dame, IN 46556, USA\\
$^{10}$Okayama Astrophysical National Astronomical Observatory, 3037-5 Honjo, Kamogata, Asakuchi, Okayama 719-0232, Japan\\
$^{11}$Nagano National College of Technology, Nagano 381-8550, Japan\\
$^{12}$Tokyo Metropolitan College of Industrial Technology, Tokyo 116-8523, Japan\\
$^{13}$School of Chemical and Physical Sciences, Victoria University, Wellington, New Zealand\\
$^{14}$University of Canterbury Mt John Observatory, PO Box 56, Lake Tekapo 7945, New Zealand\\
$^{15}$Department of Physics, Faculty of Science, Kyoto Sangyo University, 603-8555 Kyoto, Japan\\
$^{16}$Department of Astronomy, Ohio State University, 140 W. 18th Ave.,Columbus, OH~43210, USA\\
$^{17}$Department of Physics, University of Warwick, Gibbet Hill Road, Coventry, CV4~7AL, UK
}

\date{Accepted 2017 April 27. Received 2017 April 27; in original form 2017 March 24}

\pubyear{2016}

\begin{document}
\label{firstpage}
\pagerange{\pageref{firstpage}--\pageref{lastpage}}
\maketitle

\begin{abstract}
We report discovery of the lowest mass ratio exoplanet to be found by the microlensing method in the light curve of the event OGLE~2016--BLG--1195. This planet revealed itself as a small deviation from a microlensing single lens profile from an examination of the survey data. The duration of the planetary signal is $\sim 2.5\,$hours. The measured ratio of the planet mass to its host star is $q = 4.2\pm 0.7 \times10^{-5}$. We further estimate that the lens system is likely to comprise a cold $\sim$3 Earth mass planet in a $\sim\,$2 AU wide orbit around a 0.2 Solar mass star at an overall distance of 7.1 kpc.
\end{abstract}




\section{Introduction}

In the technique of gravitational microlensing, planetary systems are utilized as naturally occurring lenses of light from background source stars \citep{1991ApJ...374L..37M, 1992ApJ...396..104G, 1994ApJ...436..112B}. 
In this technique, one observes the magnification of the source star as the lens star moves across the line-of-sight from Earth. If the lens star has planets, then additional lensing can occur producing perturbations in the profile one would otherwise expect for a single lens. Interestingly, the planetary signal strength is not necessarily weaker for low mass planets, making the technique of microlensing capable of detecting planets down to Earth mass for ground based projects \citep{1996ApJ...472..660B}  and Mars mass for space based projects \citep{2009astro2010S..18B}.

Of the approximately 3500 extrasolar planets so far discovered, most have been detected by the radial velocity technique \citep{2006ApJ...646..505B} or transit technique \citep{2015ApJS..217...31M}. The radial velocity and transit techniques are most sensitive to warm planets with close-in orbits around the host stars. In contrast, gravitational microlensing is most sensitive to cold planets in wider orbits. In planetary formation, an important delimiter in the protoplanetary disk is the ``snowline'', beyond which water remains as ice during the planetary formation process \citep{1993ARA&A..31..129L, 2004ApJ...616..567I}. It is important to understand the process of planetary formation beyond the snowline, and microlensing is well suited to probe this important region of parameter space.

To date, there have been 51 published discoveries of extrasolar planets by microlensing. Most of these have estimated masses above the 12--15 Earth mass threshold that separates the low mass rocky planets from the gas giants. Statistical measures have been derived from microlensing data for giant planets
\citep{2011Natur.473..349S, 2010ApJ...720.1073G, 2016MNRAS.457.4089S, 2012Natur.481..167C} with a recent study showing a break in the power law distribution of the planet:host star mass ratio at around 
10$^{-4}$ \citep{2016ApJ...833..145S}. It is important to probe the distribution of planets with mass ratios below this value. Here we report a microlensing discovery of a planet with the lowest ratio of its mass to its host star amongst microlensing planets.

\section{Observations}

The microlensing event OGLE 2016--BLG--1195 (hereafter ob161195)  was discovered by the OGLE-IV survey and was alerted  by the Early Warning System \citep{2003AcA....53..291U} on June 27.57, 2016 (UT). The equatorial coordinates of the event are: $\alpha$=17:55:23.50 $\delta$=$-$30:12:26.1 (J2000.0). OGLE-IV monitoring of the event was conducted with the 1.3-m Warsaw telescope located at the Las Campanas Observatory, Chile. The telescope was equipped with the 32 CCD mosaic camera covering 1.4 square degrees with the resolution of 0.26$\arcsec$/pixel \citep{2015AcA....65....1U}. Observations were obtained through the standard $I$-band filter. ob161195 was located in one of the frequently observed fields with the standard cadence once per hour. Unfortunately, it occurred during the microlensing Kepler K2C9 campaign \citep{2016PASP..128l4401H} and was located outside the superstamp monitored continuously by the satellite. Thus, the OGLE cadence of this field for the time of the K2C9 campaign was reduced to three per night.

ob161195 was alerted by the MOA collaboration as MOA~2016--BLG--350 on 2016 July 28 10:55 UT approximately 20 hours after the OGLE alert. Since 2006, the MOA microlensing survey has employed a 1.8 m telescope and 80 megapixel camera \citep{2008ExA....22...51S} at the University of Canterbury Mt John Observatory near Lake Tekapo, New Zealand. During a single exposure the MOA camera captures a field of view of 2.4 square degrees, with 23 separate target fields on the sky for microlensing survey observations. MOA employs a high cadence observational strategy that aims to routinely cycle through these target fields as many times per night as possible, with some fields being observed more often than others. MOA routinely surveys the Galactic Bulge with a custom broadband red filter, hereafter denoted $\RM$, which corresponds approximately to the sum of the standard $I$ and $R$ passbands. Occasional observations (once per night) are made in the visual band filter, hereafter $\VM$, which approximates the standard $V$ passband. The $\RM$ observations are reduced in real-time as part of the analysis pipeline that is designed for detecting microlensing events and other astrophysical transients \citep{2001MNRAS.327..868B}. The $\VM$ observations are reduced offline as it is not necessary to do this in real-time in our detection of transient events.

Event ob161195 occurred in one of the MOA fields that is sampled every 16 minutes. The light curve was well sampled for several days around the peak of the event with the only interruptions due to daylight. On the second night after the MOA alert, visual inspection of observations revealed a possible microlensing anomaly in progress. The lightcurve profile featured a small perturbation that resembles what one would expect from a low mass planet in the lens system \citep{1996ApJ...472..660B}. The MOA observational cadence of the corresponding field was increased once this anomaly was noticed. An alert was issued to the microlensing community just after the peak of the perturbation but this feature was over within an hour of this alert. As a result, no effective follow-up observations could be carried out. Preliminary models were circulated that indeed showed the perturbation was likely caused by a planet orbiting the lens star. Because the ob161195 planetary anomaly was very short and occurred during Chilean day time it was therefore not possible to confirm the planetary perturbation in the OGLE data. 

This event is in the observational footprint of the new Korean Microlensing Telescope Network which operates three microlensing survey telescopes in Australia, South Africa, and Chile \citep{2016JKAS...49...37K}. Their coverage was were not influenced by the MOA and OGLE alerts. However, their data confirm the existence of this perturbation and, in the interests of a completely independent analysis, their observations are presented separately \citep{2017arXiv170308548S}.

In this work, we present the analysis of the MOA and OGLE data from the point of view of the discovery observations.

\section{Data Analysis}

\subsection{Difference Imaging Photometry}
\label{ep}

In order to obtain optimized photometry, we carried out an offline re-reduction of $\RM$ and $\VM$ images obtained by extracting sub-images centered on the event from the larger observation images. For this offline analysis we selected observations from mid 2011 to the end of 2016. Difference imaging was used to derive the photometry, with each of the MOA passbands treated separately with their own reference images. For offline analysis, we use our own implementation that incorporates a numerical kernel as described by \citet{2008MNRAS.386L..77B} with our own modification to allow for a spatial variation of the kernel across the field-of-view in a similar manner to that given by \cite{2000A&AS..144..363A}.

Microlensing events are observed in crowded fields and their centroids on the images are often blended with neighbouring stars. It is important that these centroids are measured carefully because the nearest resolved star on the image will not necessarily be the source star for the microlensing event. The MOA difference images are measured using a reference image analytical PSF model of the form used in the Dophot photometry code of \citet{1993PASP..105.1342S}. This PSF is then used to measure a given difference image after convolving with the kernel for that observation image. The model PSF is optimized by finding the centroid and shape parameters that give the best photometry on a set of difference images where the flux of the source star is significantly magnified.

The baseline photometry, in $\RM$ and $\VM$ separately, was examined during the off-event years (ie 2011--2015) for any correlations that may be present due to variations in the seeing and the effects of differential refraction (parameterized by the hour angle). We find some small effects present and so we ``detrend'' the data by modelling the baseline as a function of seeing and the hour angle. We see an improvement in the standard $\chi^2$ goodness of fit of $\Delta\chi^2\approx698$ in $\RM$ and $\Delta\chi^2\approx23$ for $\VM$ in the baseline. These respective $\RM$ and $\VM$ models were used as corrections to photometry which were subtracted off all the data including those where the source is magnified. Furthermore, all observations with FWHM worse that 4.5 pixels were rejected. This resulted in 13969 $\RM$ band and 253 $\VM$ band measurements that are used in this study. The OGLE data completed our photometry set with 5365 measurements.

Difference imaging photometry measures flux differences for the observation images with respect to the reference image. It is desirable to place these measurements onto an instrumental magnitude scale. The Dophot photometry software \citep{1993PASP..105.1342S} was run on the difference imaging reference images for each of the $\RM$ and $\RM$ passbands. The resulting list of extracted stellar objects were cross referenced with each other to produce a single catalog of field stars where instrumental magnitudes in both $\RM$ and $\VM$ could be obtained. We add to this catalog an object corresponding to the optimized centroid location (from above) of the source star for the microlensing event. The fluxes of these catalog stars were then measured on a selection of \emph{unsubtracted} $\RM$ and $\VM$ observation images using the same procedure that is used to measure the difference images, but using the PSF parameters derived from running Dophot. Linear regression was used to register these fluxes to those as measured by Dophot. Using regression again and the event photometry from the unsubtracted images as a template, the flux differences resulting from the difference imaging analysis can then be transformed onto the same flux scale as those in the Dophot $\RM$ and $\VM$ catalog.

The MOA instrumental magnitudes were calibrated by cross referencing stars in our Dophot catalog to stars in the OGLE-III catalog which provides measurements in the standard Kron-Cousins I and Johnson V passbands \citep{2011AcA....61...83S}. From this we derived the following relation between the MOA instrumental magnitudes and colours and the standard magnitudes and colours.
\begin{align*}
I_\mathrm{OGLE-III} - \RM & = [28.126\pm0.003] - [0.218\pm0.002] C_\mathrm{MOA}\\
(V-I)|_\mathrm{OGLE-III}  & = [0.505\pm0.004] + [1.105\pm0.003] C_\mathrm{MOA}
\end{align*}
where $C_\mathrm{MOA} = \VM - \RM$. With these relations, together with our catalog of Dophot measured stars, we identified the well known ``red clump giants'' on the $I_\mathrm{OGLE-III}$ vs $(V-I)|_\mathrm{OGLE-III}$ colour-magnitude diagrams. Using only stars within 2$\arcmin$ of the event position, we measured the centroid of the clump to be:
\begin{align*}
I_\mathrm{clump} &= 16.212 \pm 0.018     \\
(V-I)|_\mathrm{clump} &= 2.468 \pm 0.007 \\
\end{align*}
Adopting the intrinsic red clump colour of $(V-I)|_\mathrm{RCG,0}=1.06$ \citep{2011A&A...533A.134B} and intrinsic magnitude $I_\mathrm{RCG,0}=14.45$ \citep{2013ApJ...769...88N}, we derive the following for the extinction and reddening towards the direction of the event:
\begin{align*}
A_I &= 1.762 \pm 0.018\\
E(V-I) &= 1.408 \pm 0.007
\end{align*}

We will use these values in our subsequent modelling and analysis of the source star properties.

\subsection{Modelling the Event}
\label{me}

We modelled the light curve photometry for this event using the image-centered ray-shooting 
method \citep{1996ApJ...472..660B,2010ApJ...716.1408B} to calculate finite source effects.
This method has been tested extensively for mass ratios down to $10^{-7}$
\citep{2002ApJ...574..985B}.

\begin{figure}
\centering\includegraphics[scale=0.4]{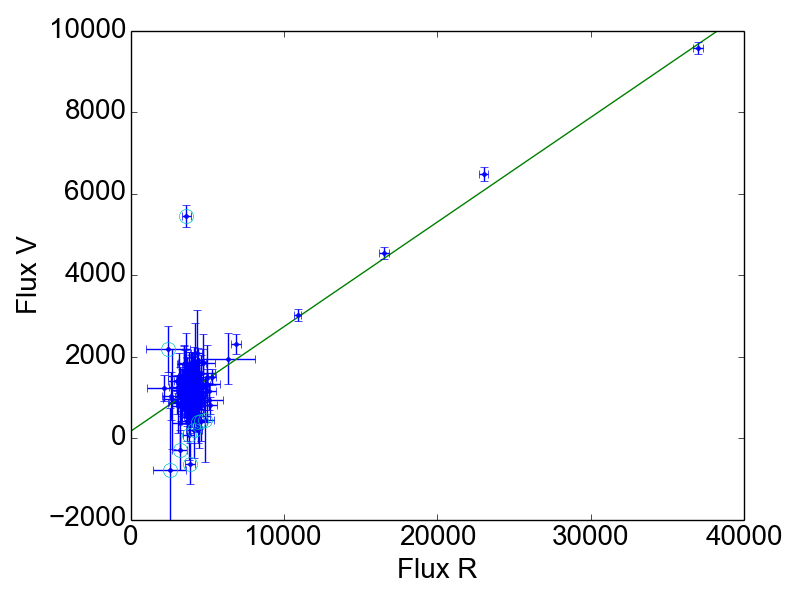}
\caption{Near simultaneous instrumental MOA $\RM$ and $\VM$ photometry measurements in linear flux units. In the data points plotted here, the difference between the $\RM$ and $\VM$ measurements is typically around 20 minutes. The slope gives the instrumental source star flux ratio which in turn gives the instrumental colour from which the calibrated $V-I$ colour can be derived. The circled data points are outliers that were iteratively removed from the fitting procedure.}
\label{fig:colourRV}
\end{figure}

The calculation of finite source effects requires an appropriate limb darkening model for the source star. A model independent measurement of the source star colour can obtained by plotting near simultaneous $\RM$ and $\VM$ measurements against each other as shown in Fig.~\ref{fig:colourRV}. The slope of the plot gives the ratio of source star fluxes in the respective passbands, or equivalently, the magnitude difference. We derive a model independent instrumental colour index as $\VM-\RM=1.476\pm0.029$. Using our instrumental magnitude calibrations and reddening measurements from the previous section, we obtain a dereddened colour index of $V-I=0.728\pm0.033$ for the source star. This corresponds to an effective temperature $T_\mathrm{eff}\sim6000$K \citep{1998A&A...333..231B}. In our modelling we use a linear limb darkening law with parameters appropriate to this value of $T_\mathrm{eff}$ and metallicity $\log g=4.5$.

Our magnification profile for a binary lensing model is described by the following parameters: the time of closest approach to the barycentre $t_0$; the Einstein radius crossing time $t_\mathrm{E}$; the impact parameter, $u_0$, in units of the Einstein ring radius of the source star trajectory with respect the binary lens barycentre; the ratio $q$ of the secondary lens component to the primary; the source radius crossing time $t_*$; the separation, $s$, of the binary lens components projected onto the a plane at the lens system perpendicular to the Earth-source line-of-sight; and the angle, $\phi$, the source star trajectory makes with the planet-star separation.

In our modelling we sought an optimal set of these parameters that can jointly model our observations in the MOA $\RM$ and $\VM$ data and the OGLE data. The observational data are fully described by the 7 parameters that describe the magnification profile together with 2 flux scaling parameters for each of the passbands. Our goodness of fit is assessed using the standard $\chi^2$ measure combined from the three passbands we use here. We employed the standard technique of searching the phase space of 7 parameters that describe the magnification profile using the Markov Chain Monte Carlo method to find those parameters that minimize the value of $\chi^2$. The measured binary lens model parameters so derived are listed in Table~\ref{tab:params}. The uncertainties in the parameters correspond to their respective range of values that satisfy the standard criterion of $\chi^2<\chi_\mathrm{min}^2+1$.

\begin{table}
\begin{tabular}{lll}
Parameter & wide model & close model\\
\hline
$t_\mathrm{E}$ /days   & $10.16 \pm 0.25$        & $10.21 \pm 0.26$ \\
$t_0$ / HJD$-$2450000  & $7568.7719 \pm 0.0020$   & $7568.7713\pm0.0020$\\
$u_0$                & $0.0514\pm0.0014 $       & $0.0512\pm0.0015$\\
$q$ /$10^{-5}$        & $4.25 \pm 0.67$          & $4.20 \pm 0.65$\\
$t_*$ / days         & $0.0336 \pm 0.0023$      & $0.03379 \pm 0.0021$\\
$s$                  & $1.0698\pm0.0078$        & $0.99570\pm0.0073$\\
$\phi$               & $55\fdg31\pm0.26$        & $55\fdg26\pm0.26$ \\
$\chi_\mathrm{min}^2$ (19587 data points)           & 19580.4 & 19581.4
\end{tabular}
\caption{Best fitting binary microlensing model parameters.}
\label{tab:params}
\end{table}

\begin{figure}
\centering\includegraphics[scale=0.35]{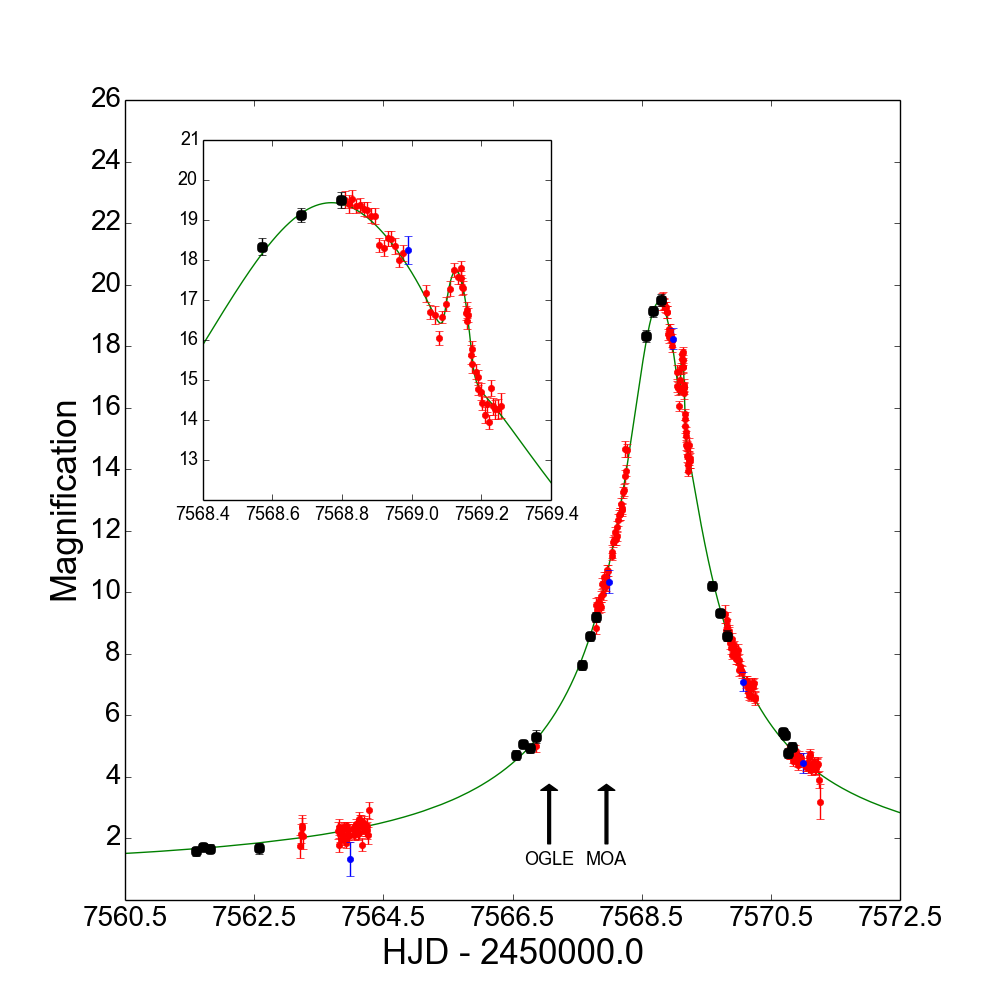}
\caption{Lightcurve showing photometry and modelling of ob161195. The photometry are normalized to the magnification values according to the best fitting planetary microlensing model shown by the green curve. The data points shown are MOA $\RM$ (red), MOA $\VM$ (blue), and OGLE (black). The main figure plots the data over a 12 day period where the planetary deviation can be seen in relation to the data on other nights. The inset shows a close up of the deviation. The arrows mark the times of the MOA and OGLE alert notifications.}
\label{fig:lc}
\end{figure}

We find two possible solutions: a ``close'' model and a ``wide'' model where the projected separation onto the lens plane is either inside or outside the Einstein radius. The lightcurve  together with the best fitting wide model is plotted in Fig~\ref{fig:lc}. The wide model is only slightly favoured at an insignificant level of $\Delta\chi^2\approx1$. Formally, we are unable to distinguish between the two models. As expected, we find the observed lightcurve is best reproduced with a binary lens model of extreme mass ratio corresponding to a planetary mass for the secondary. The very small measured mass ratio below $10^{-4}$, for the close and wide models, is striking here. The planetary perturbation is covered only by the MOA $\RM$ data and is the dominant contributor to measurement of the planetary microlensing parameters. The addition of the OGLE data allows for a tighter constraint on our measurement of $t_\mathrm{E}$.

\begin{figure}
\centering\includegraphics[scale=0.45]{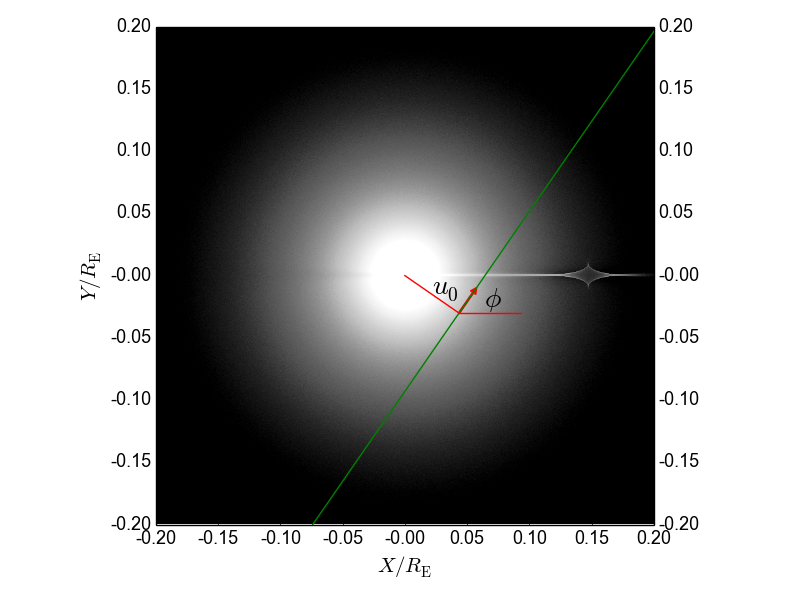}
\centering\includegraphics[scale=0.45]{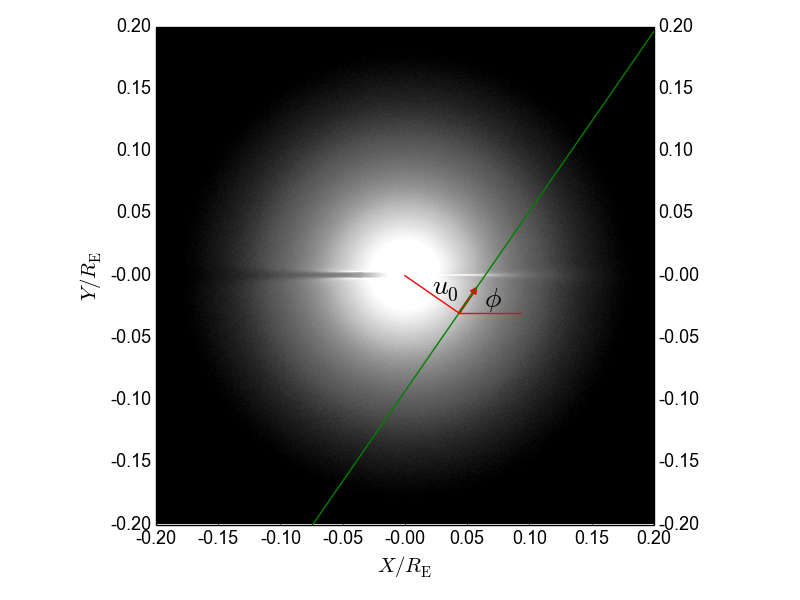}
\caption{Magnification maps for the wide (upper panel) and close (lower panel) binary microlensing models. The green line shows the source star track across the maps. The magnification at a given position on the map is the integration over the source star size (smaller than the pixel scale here).}
\label{fig:magmap}
\end{figure}

Magnification maps are a useful tool for visualizing the possible magnification profiles for a given source star trajectory in a microlensing event \citep{1997MNRAS.284..172W}. In Fig.~\ref{fig:magmap}, we show these maps for the close and wide solutions. The high sensitivity in this event to such a low mass ratio planetary system is because the planet is very close to the Einstein ring in both the close and wide cases. This results in a region of enhanced magnification on the map that extends a long way along the line separating the planet and the lens star. The planetary perturbation occurs when the source star crosses this line. For this lens system geometry, a wide range of possible values of $u_0$ would have resulted in a planetary perturbation if one had been observing at that time. We note that the central time and width of the deviation is consistent with that in a relation given by \citet{2013MNRAS.431.2975A} who study this effect for events with high peak magnifications where $u_0<0.02$.

Due to the short timescale of the event, we could not measure the microlensing parallax effect. 

It is a possible that a short term planetary perturbation could be mimicked by a binary source of extreme flux ratio where the fainter companion gets highly magnified \citep{1998ApJ...506..533G}. We attempted to model our observations with a static binary source single lens model. We parameterize this model with the Einstein crossing time, $t_\mathrm{E}$ of the binary source together with the time of closest approach, $t_0$, and impact parameter, $u_0$ of the primary component. We then introduce five additional parameters for the secondary companion of the primary. These are its dimensionless separation, $d$, from the primary, its position angle, $\psi$, with respect to the trajectory of the primary, and its flux ratio, $\alpha$, to the primary, the ratio $\rho$ of the angular size of the companion to that of the Einstein ring, and the coefficient, $\lambda$, of a linear limb darkening law. We consider finite source effects for the secondary because, if the binary source model is to account for the perturbation we observe, the secondary is expected to be highly magnified and pass close to the lens.

\begin{figure}
\centering\includegraphics[scale=0.45]{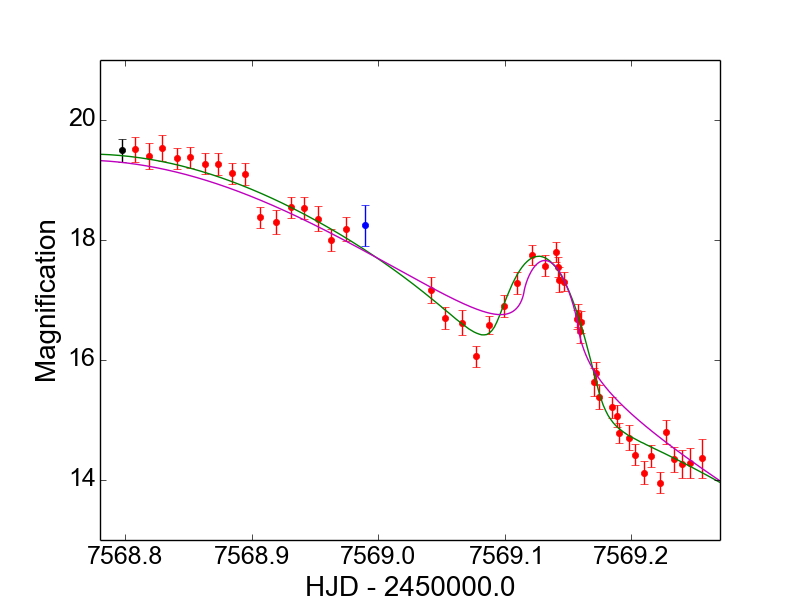}
\caption{Close up of the observed perturbation comparing the best fitting binary source model (magenta) with the best fitting wide planetary model (green). As before the data points are $\RM$ (red), $\VM$ (blue), and OGLE (black).}
\label{fig:plbs}
\end{figure}

The best fitting parameter values are listed in Table~\ref{tab:bs}. The negative value of $u_0$ together with the values for $d$, $\psi$, and $\rho$ mean that the secondary lags behind the primary and passes over the lens over the course of the event. With a flux ratio of $\sim$0.002, the secondary is significantly fainter than the primary but is more highly magnified as expected. In Fig.~\ref{fig:plbs} we show a close up view of the observed perturbation together with the best fitting binary source model and the wide planetary microlensing model. Overall, the planetary model does a better job at reproducing the features of this perturbation. In particular, the binary source model does not fit the beginning and end of the perturbation as well as the planetary model. The difference in the goodness-of-fit between the two models is $\Delta\chi^2\approx120$. We can compare this to the similar case of OGLE~2005--BLG--390 where a binary source model was excluded in favour of a planetary model at $\Delta\chi^2\approx46$ \citep{2006Natur.439..437B}.

The parameters of planetary and binary source models considered here are not nested parameters. Strictly speaking, the difference in $\chi^2$ is not an appropriate measure to compare the two. Following the approach of \citet{2016ApJ...825..112S}, we compare the models using Akaike's Information Criterion $\mathrm{AIC}=\chi^2+n_\mathrm{param}$ and the Bayesian Information Criterion $\mathrm{BIC}=\chi^2+n_\mathrm{param} \ln(N_\mathrm{data})$. These are standard criteria used to select a preferred model and they penalize for the number of parameters used. In our data, the planetary model gives the smaller value for both of these criteria. We find for the difference between the models: $\Delta\mathrm{AIC}\approx121$ and $\Delta\mathrm{BIC}\approx129$.  Here, we conclude that the binary source model is excluded in favour of the planetary models for ob161195.

\begin{table}
\begin{tabular}{ll}
Parameter & value\\
\hline
$t_\mathrm{E}$ /days      & $10.55\pm0.11$ \\
$t_0$ / HJD$-$2450000    & $7568.76155\pm0.00080$ \\
$u_0$                    & $-0.04945\pm0.00060$ \\
$d$                      & $0.0609\pm0.0011$ \\
$\psi$                   & $125\fdg170\pm0.015$\\
$\alpha$                 & $0.00199\pm0.00010$\\
$\rho$                   & $0.00204\pm0.00010$\\
$\lambda$                & $0.3781\pm0.0017$\\
$\chi_\mathrm{min}^2$ (19587 data points) & 19699.7
\end{tabular}
\caption{Best fitting parameters for the binary source model.}
\label{tab:bs}
\end{table}

\subsection{Observed Source Star Properties}

The source star fluxes are measured as scaling parameters when determining the best fitting microlensing magnification profile. Though sparsely sampled, the MOA $\VM$ measurements cover parts of the light curve where the source star is magnified allowing a measurement of the source star flux in this passband. For the MOA data, we derive an instrumental source star magnitude of $\RM=-8.226\pm0.001$ and an instrumental colour $\VM-\RM=1.457\pm0.018$. The MOA instrumental colour is in good agreement with the model independent value presented in the previous section. Using our instrumental calibration from Section~\ref{ep}, the apparent source star magnitude and colour in the OGLE-III system is
\begin{align*}
I_\mathrm{src} &= 19.581\pm0.001\\
(V-I)|_\mathrm{src} &= 2.113\pm0.020\\
\end{align*}
From our measurements of the red clump in Section~\ref{ep}, the extinction corrected and dereddened source star magnitude and colour is
\begin{align*}
I_\mathrm{src,0} &= 17.819\pm0.018\\
(V-I)|_\mathrm{src,0} &= 0.705\pm0.022\\
\end{align*}

\begin{figure}
\centering\includegraphics[scale=0.4]{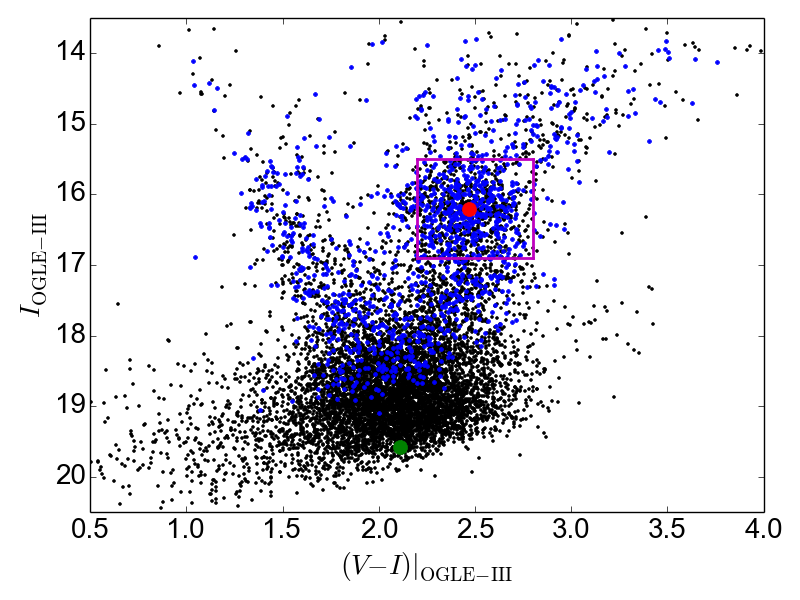}
\caption{Magnitudes and colours measured for resolved stars within $2\arcmin$ of the event ob161195. The measurements plotted in blue are instrumental magnitudes extracted from the MOA $\RM$ and $\VM$ reference images that have been calibrated to the OGLE-III $I$ and $V$ system. The data points plotted in black are the $V-I$ and $I$ values taken directly from the OGLE-III catalog. The magenta coloured box shows those stars used to measure the red clump giant centroid shown by the red point. The green point shows the measured magnitude and colour of the event source star as derived from the best fitting wide binary model described in the text.}
\label{fig:cmd}
\end{figure}

An OGLE independent determination of the dereddened color based on OGLE-IV photometry yields $(V-I)_0 = 0.67\pm0.03$. This is consistent with
MOA result. In Fig.~\ref{fig:cmd} we plot a $I_\mathrm{OGLE-III}$ vs $(V-I)_\mathrm{OGLE-III}$ colour magnitude diagram for MOA and OGLE measurements of resolved stars together with the above source star measurements. The position of the source star magnitude and colour measurements is well below the red clump and sub giant regions. Main sequence stars could not be resolved in our Dophot measurements of the reference images. However, we can compare our $(V-I, I)|_\mathrm{src,0}$ measurement with the colour magnitude diagram of main sequence stars of \citet{1998AJ....115.1946H} based on HST observations of Baade's window. After allowing for extinction and reddening based on the red clump measurements of the HST data \citep{2008ApJ...684..663B}, our measurements of the source star magnitude and colour are placed well within the main sequence star distribution. Using Table~1 of \citet{1998A&A...333..231B} for $\log g=4.5$, our value for $(V-I)|_\mathrm{src,0}$ implies an effective temperature of 5820 K---not too dissimilar to value of 5770 K for the Sun. We conclude that the source star in this event is a Solar-like star.

These measurements can be used to calculate the angular source star radius, $\theta_\mathrm{src}$ using the follow relation involving the apparent magnitude and colours in the OGLE-III system \citep{2014AJ....147...47B, 2015ApJ...809...74F}
\begin{equation*}
\log 2\theta_\mathrm{src} = 0.5014 + 0.4197 (V-I) - 0.2I
\end{equation*}
This gives $\theta_\mathrm{src}=0.856\pm0.019$ $\mu$as.

\subsection{Estimation of Lens System Parameters}

For a number of events, it is possible to detect microlensing parallax as the Earth moves on its orbit, and then derive measurements of the absolute masses of the lens system components together with the distance to the lens. The microlensing Einstein ring crossing timescale of around 10 days is too small to allow a detection of parallax. Following the approach of \citep{2008ApJ...684..663B} we employ standard Bayesian techniques to estimate the parameters in the lensing system. To derive distributions of lens distances and velocities we use a Galactic model comprising a double-exponential disk \citep{2002AJ....124.2721R} and a bar model from \citep{1995ApJ...447...53H}. We constrain the mass of the lens and its distance to one relation by a measurement of the angular radius of the Einstein ring. From the microlensing parameters in Section~2 and the source star angular size measurement from Section~\ref{me}, we have
\begin{equation*}
\theta_\mathrm{E} = \theta_\mathrm{src} \frac{t_\mathrm{E}}{t_*} = 0.261\pm0.020~\mathrm{mas}
\end{equation*}

We make the assumption that planets are equally likely around stars regardless of the planet mass, and the mass of the host star and its distance. We further assume that the  planetary orbital planes have random and uniform orientations. In Table~\ref{tab:bayes}, we list our estimations of the planet and host star mass, the 3D orbital separation, and the distance to the planetary system. As the characteristic measurement we take the value that maximizes its respective likelihood function. Also provided are the upper and lower limits at the 68\% and 99.7\% confidence levels. We see very little difference in the estimated values of the parameters when comparing the close and wide models. Even the values for the orbital separation are effectively in agreement. Qualitatively, we have a planet at around $\sim$1.8 AU from its host star that could be just within or just outside the Einstein ring. In Table~\ref{tab:bayes}, we present the averages of their respective close and wide values.

\begin{table}
\begin{tabular}{p{1.4cm}p{0.6cm}p{0.6cm}p{0.5cm}p{0.5cm}p{0.5cm}p{0.5cm}p{0.5cm}}
Parameter & Mode & Median & \multicolumn{2}{c}{68\% Limits} & \multicolumn{2}{c}{99.7\% Limits} & Units\\
& & & lower & upper & lower & upper & \\
\hline
$M_\mathrm{planet}$      & 2.74 & 5.10 & 2.25 & 10.6 & 0.25 & 22.3 & $M_\mathrm{Earth}$\\
$M_\mathrm{host }$       & 0.19 & 0.37 &  0.16 & 0.75 & 0.02 & 1.29 & $M_\odot$\\
$a_\mathrm{3D}$ (close)  & 1.77 & 1.98 & 1.60 & 3.02 & 0.57 & 12.4 & AU\\
$a_\mathrm{3D}$ (wide)   & 1.93 & 2.15 & 1.73 & 3.28 & 0.61 & 13.6 & AU\\
$D_\mathrm{lens}$        & 7.27 & 7.20 & 6.18 & 8.05 & 2.01 & 9.71 & kpc
\end{tabular}
\caption{Estimated lens parameters derived from a Bayesian analysis. The mode is the value of the parameter that maximizes the value of the associated likelihood function. The median is the value that divides the likelihood function into two parts of equal area. The upper and lower limits correspond to 68\% and 99.7\% confidence levels.}
\label{tab:bayes}
\end{table}

\section{Discussion}

The planet ob161195Lb has the lowest mass ratio, measured to date, amongst microlensing planets that orbit a single star without any binary companions. The closest contender is the planet in the binary system OGLE-2013-BLG-0341L \citep{2014Sci...345...46G}, assuming that the wide binary model, rather than the circumbinary model, is correct for that event. This planet is the sixth microlens planet with a sub-$10^{-4}$ mass ratio---the others being OGLE-2013-BLG-0341LBb, OGLE~2005--BLG--390Lb \citep{2006Natur.439..437B},
OGLE~2005--BLG--169Lb \citep{2006ApJ...644L..37G, 2015ApJ...808..169B,2015ApJ...808..170B}, 
OGLE-2007-BLG-368Lb \citep{2010ApJ...710.1641S}, and
MOA~2009--BLG-266Lb \citep{2011ApJ...741...22M}.

Because the planetary mass ratio is measured in all planetary microlensing detections,
the statistical properties of the planetary systems probed by microlensing are most easily
described in terms of the mass ratio. A recent statistical analysis of the planetary
signals detected by the MOA survey from 2007 through 2012 \citep{2016ApJ...833..145S} 
was able to identify a power law break in the mass ratio function based on 23 planets from 
the MOA survey and 7 additional planets from previous analyses 
\citep{2010ApJ...720.1073G,2012Natur.481..167C}. Due to the lack of detected planets with mass
ratios lower than $q = 3\times 10^{-5}$, the precise location of this break is somewhat
uncertain, at $q_{\rm break} = 6.7{+9.0\atop -1.8}\times 10^{-5}$. Thus, the newly
discovered planet, ob161195Lb, is close to the mass ratio function break, and
it does not provide a strong constraint on the behavior of the mass ratio function below
the break. It is likely to be an example of the most common type of planet that orbits
beyond the snow line.

Although the MOA observational cadence was increased just after the planetary perturbation was noticed, this planet can be regarded as a near ``blind survey only'' detection where the planetary signal was noticed after the fact. This contrasts with the classic ``followup mode'' where a real-time alert of either a high magnification or planetary perturbation in progress results in subsequent followup observations by other telescopes or the survey telescope itself. It is expected that most planetary discoveries in new generation microlensing projects KMTNet \citep{2016JKAS...49...37K} and WFIRST \citep{2009astro2010S..18B} will be of the blind survey type. Previous examples of these types of microlensing planet detections are the giant planets MOA~2011--BLG--322Lb \citep{2014MNRAS.439..604S}, MOA~2015--BLG--353Lb \citep{2015MNRAS.454..946R}, and OGLE~2012--BLG--0950Lb \citep{2017AJ....153....1K} and the rocky planet MOA~2007--BLG--192 \citep{2008ApJ...684..663B}.

Our statistical analysis in estimating the absolute parameters of this planetary system, does not rule out the possibility of a super Earth planet orbiting a Solar-like star just within the outer edge of the liquid water habitable zone. For the proposed WFIRST microlensing survey, it is expected that a small, but not insignificant, fraction of the microlensing planet yield will comprise planets in the habitable zone \citep{2009astro2010S..18B}. High resolution follow-up observations by Keck reveal a possible supermassive planet within the habitable zone of the low mass star star in MOA~2011--BLG--293 \citep{2014ApJ...780...54B}.

However, it is more likely that ob161195Lb is a cold rocky super Earth orbiting an M star. Qualitatively, there is little difference between the close and wide solutions for the 3D orbital separation. Both place the planet at $\sim$2 AU which would place the planet beyond the snowline, and habitable zone, of its M-star host. It is worth noting that stellar insolation is not the only source of heating for a planet. Internal heating, tidal friction \citep{2010RPPh...73a6901B}, volcanism \citep{2010AJ....140.1370K}, and radiogenic heating \citep{2014Icar..243..274F} can also contribute to the heat budget of a planet to allow liquid water below the planet's surface. The \emph{extended} sub-surface habitable zones could be more than ten times larger than the circumstellar habitable zones of stars \citep{2013P&SS...85..312M}. It is possible that ob161195Lb is a rocky planet in this extended habitable zone.

It is therefore important to determine the absolute values for planet mass, host star mass, and orbital separation in ob161195L. Unfortunately, this event lies outside the fields-of-view of the recent K2C9 survey \citep{2016PASP..128l4401H} so a parallax measurement combining ground based and Kepler observations is not possible here. The 95\% upper limit on the mass estimation of the lens star in 1.1$M_\odot$. This corresponds to an upper limit on the brightness of the lens star of $I\gtrsim19.5$. The best prospects for measuring the physical properties of the planet would be to use follow-up high resolution imaging.

\section{Conclusions}

We have discovered a low mass planet in the microlensing event ob161195. This planet has the lowest mass ratio and lowest mass fraction amongst microlensing planets so far detected. Although the absolute masses of the lens system components could not be measured, the measured mass ratio here is an important additional data point in the so far limited sample of mass ratio measurements below the newly discovered break at $q\sim10^{-4}$.

\section*{Acknowledgements}

The MOA project is supported by JSPS Kakenhi grants JP24253004, JP26247023, JP16H06287, JP23340064 and JP15H00781 and by the Royal Society of New Zealand Marsden Grant MAU1104. The OGLE project has received funding from the National Science Centre, Poland, grant MAESTRO 2014/14/A/ST9/00121 to AU. NJR is a Royal Society of New Zealand Rutherford Discovery Fellow. AS is a University of Auckland Doctoral Scholar.




\bibliographystyle{mnras}
\bibliography{mybibfile} 






\bsp	
\label{lastpage}
\end{document}